# Observation of E$_8$ Particles in an Ising Chain Antiferromagnet

Zhao Zhang[1], Kirill Amelin[2], Xiao Wang[1], Haiyuan Zou[1], Jiahao Yang[1], Urmas Nagel[2], Toomas Rõõm[2], Tusharkanti Dey[3,4], Agustinus Agung Nugroho[5], Thomas Lorenz[3], Jianda Wu[1]*, Zhe Wang[3]†

[1]Tsung-Dao Lee Institute and School of Physics & Astronomy, Shanghai Jiao Tong University, Shanghai 200240, China
[2]National Institute of Chemical Physics and Biophysics, 12618 Tallinn, Estonia
[3]Institute of Physics II, University of Cologne, 50937 Cologne, Germany
[4]Department of Physics, Indian Institute of Technology (Indian School of Mines), Dhanbad 826004, Jharkhand, India
[5]Faculty of Mathematics and Natural Sciences, Institut Teknologi Bandung, 40132 Bandung, Indonesia

Near the transverse-field induced quantum critical point of the Ising chain, an exotic dynamic spectrum consisting of exactly eight particles was predicted, which is uniquely described by an emergent quantum integrable field theory with the symmetry of the E$_8$ Lie algebra, but rarely explored experimentally. Here we use high-resolution terahertz spectroscopy to resolve quantum spin dynamics of the quasi-one-dimensional Ising antiferromagnet BaCo$_2$V$_2$O$_8$ in an applied transverse field. By comparing to an analytical calculation of the dynamical spin correlations, we identify E$_8$ particles as well as their two-particle excitations.

Exotic states of matter, such as high-temperature superconductivity or magnonic Bose-Einstein condensation, can emerge in the vicinity of a quantum critical point [1], which identifies a zero-temperature phase transition tuned by an external parameter, e.g. chemical substitution or applied magnetic field [2,3]. Quantum critical points are often characterized by enhanced many-body fluctuations together with divergence of correlation length and complex emergent symmetry [1,4,5,6,7,8], thus it is generally a formidable task to precisely describe the quantum many-body physics near a quantum critical point. Exactly solvable models play a crucial role in this regard, because a precise understanding of the quantum many-body physics can be gained by rigorously analyzing these models [4,6]. The one-dimensional (1D) spin-1/2 Ising model in a transverse magnetic field is such a paradigmatic example [1,4,5,6,7,8,9]. Considering only the exchange interaction between the nearest-neighbor spins on a chain [10,11], this model has been investigated most broadly in quantum magnetism, which provides deep insights into the fundamental aspects of the quantum many-body physics [1,6,7,8]. In particular, highly unconventional dynamic properties have been theoretically predicted to emerge near the transverse-field Ising quantum critical point, either for equilibrium states upon constant perturbations or for states far from equilibrium after a quantum quench (see e.g. Refs. [12,13,14,15,16,17,18]). Moreover, the study of the transverse-field Ising quantum critical point is of importance also in the context of quantum information [5,8] and quantum simulation using ultracold atoms [19].

A remarkable prediction of an exotic dynamic spectrum was made three decades ago for the transverse-field Ising chain perturbed by a small longitudinal field [12]. It is described by the Hamiltonian

$$H = -J \sum_i S_i^z S_{i+1}^z - B_\perp \sum_i S_i^x - B_\| \sum_i S_i^z \quad (1)$$

with the $x$ and $z$ components $S_i^x$ and $S_i^z$, respectively, of the spin-1/2 magnetic moment at the $i^{\text{th}}$ site on a 1D chain. The first term is the Ising term with the ferromagnetic exchange $J > 0$ between the nearest-neighbor spins. The second and third terms describe the interactions of the spins with the transverse field $B_\perp$ and the perturbative longitudinal field $B_\|$, respectively. Close to the transverse-field Ising quantum critical point [see Fig. 1(b)], the excitation spectrum of this model was predicted to be governed by a complex symmetry which is described by a quantum integrable field theory with the E$_8$ symmetry (an exceptional simple Lie algebra of rank 8) [12], which, however, is rarely explored experimentally. An analytical solution of the E$_8$ excitation spectrum delivered exactly eight particles ($m_1$ to $m_8$), the existence of which is uniquely determined by the specific ratios of their masses (Table 1) with the lowest mass scaling with the perturbative longitudinal field, i.e. $m_1 \propto |B_\||^{8/15}$ [12]. Further analysis on the dynamic characteristics of the eight particles showed that the single-particle spectral weight decreases monotonically and drastically with increasing energy [Fig. 1(a)] [13,14]. Despite the apparent simplicity of the spin Hamiltonian in Eq. (1), an experimental realization of the E$_8$ spectrum, however, is very difficult, because several crucial criteria must be simultaneously fulfilled: one-dimensionality of spin interactions, strong Ising anisotropy, and a perturbative longitudinal field.

In this Letter, we use high-resolution terahertz (THz) spectroscopy to resolve E$_8$ particles in an antiferromagnetic Ising spin-chain material BaCo$_2$V$_2$O$_8$, where all the crucial criteria are found to be realized. By performing analytical calculation of spin dynamic structure factor using the quantum integrable field theory of the E$_8$ spectrum, we unambiguously identify E$_8$ single particles as well as their two-particle excitations.

BaCo$_2$V$_2$O$_8$ is a magnetic insulator with a tetragonal crystal structure [20,21]. Based on the magnetic cobalt ions, the spin chains in BaCo$_2$V$_2$O$_8$ are constituted by edge-sharing CoO$_6$ octahedra, running with a four-fold screw axis along the crystallographic $c$-axis [Fig. 1(c)]. High-quality single crystals of BaCo$_2$V$_2$O$_8$ were grown using the floating-zone method [21]. The crystal structure and magnetic properties were characterized by X-ray diffraction, magnetization, heat-capacity, and dilatometry measurements [21]. For the optical experiment, single crystals were oriented at room temperature using X-ray Laue diffraction and cut perpendicular to the tetragonal $a$-axis with a typical surface area of $4 \times 4$ mm$^2$ and a thickness of 0.76 mm. Using a Sciencetech SPS200 Martin-Puplett type spectrometer with a 0.3 K bolometer, THz transmission measurements were carried out down to 2.7 K (below $T_N \approx 5.5$ K) in a cryostat equipped with a superconducting magnet for applying fields up to 17 T. An external field $B_\perp$ was applied parallel to the tetragonal $a$-axis, while the THz electromagnetic waves propagated along the other tetragonal $a$-axis in Voigt configuration. A rotatable polarizer was placed in front of the sample for tuning polarization of the THz waves. The change of absorption coefficient $\Delta\alpha$ due to magnetic excitations was derived by taking the zero-field transmission spectrum at 10 K (slightly above $T_N$) as reference [33].

TABLE 1. Analytically predicted mass ratios of the $E_8$ particles ($m_1$ to $m_8$) and
the derived onsets of the multi-particle continua ($2m_1$, $m_1+m_2$, $m_1+m_3$, $3m_1$ and $2m_2$) [12,13,14].

| single | $m_1$ | $m_2$ | | $m_3$ | | $m_4$ | | $m_5$ | | | $m_6$ | | $m_7$ | $m_8$ |
|---|---|---|---|---|---|---|---|---|---|---|---|---|---|---|
| multi | | | $2m_1$ | | | | $m_1+m_2$ | | $m_1+m_3$ | $3m_1$ | | $2m_2$ | | |
| / $m_1$ | 1 | 1.618 | 1.989 | 2 | 2.405 | 2.618 | 2.956 | 2.989 | 3 | 3.218 | 3.236 | 3.891 | 4.783 |

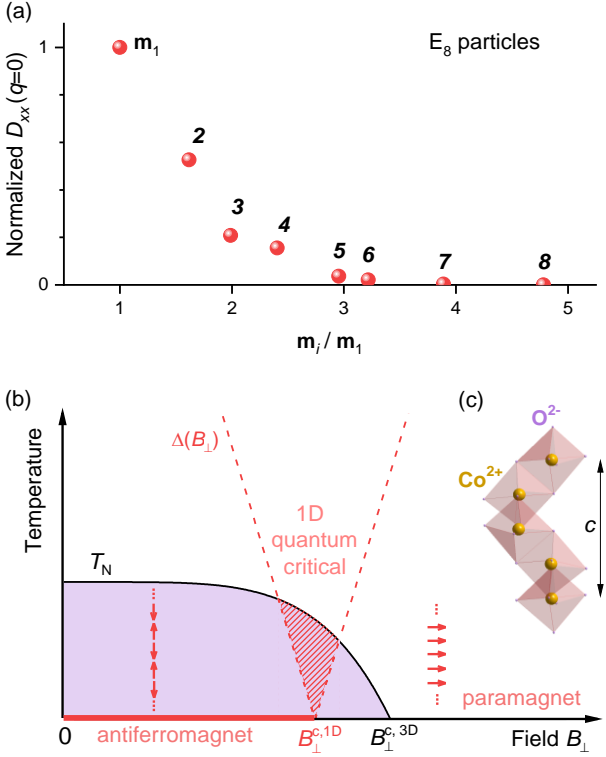

FIG. 1. (a) Normalized dynamical structure factor $D_{xx}(m_i, q = 0)$ at zero momentum transfer for the eight particles with specific ratios $m_i/m_1$ ($i = 1, 2, ..., 8$) [see Table 1 and Eq. (2)]. (b) Illustrative phase diagram of a quasi-one-dimensional Ising antiferromagnet in a transverse magnetic field. In zero field a three-dimensional (3D) Néel order is stabilized below $T_N$ due to perturbative inter-chain couplings. The quantum critical point of the transverse-field Ising chain at $B_\perp^{c,1D}$ [corresponding to the vanishing spin gap $\Delta(B_\perp)$] is masked under the 3D Néel order ($B_\perp^{c,1D} < B_\perp^{c,3D}$). A paramagnetic phase is reached when the long-range order is suppressed by $B_\perp > B_\perp^{c,3D}$. The possible region to realize the $E_8$ dynamic spectrum is indicated by the dashed area. (c) The spin chain in the quasi-one-dimensional Ising antiferromagnet $BaCo_2V_2O_8$ is constituted by edge-sharing $CoO_6$ octahedra, running with a four-fold screw axis along the crystallographic $c$-axis. In $BaCo_2V_2O_8$, the 3D Néel order is formed below $T_N \approx 5.5$ K and $B_\perp^{c,3D} = 10$ T [21].

An easy-axis anisotropy along the $c$-axis in $BaCo_2V_2O_8$ was evidenced by magnetization measurements [21], and further confirmed by investigations of quantum spin dynamics [22,23,24]. By precisely comparing to the exact results of Bethe Ansatz, the quantum spin dynamics in $BaCo_2V_2O_8$ can be nicely described by a 1D spin-1/2 antiferromagnetic Heisenberg-Ising model with a strong Ising anisotropy [22,23]. Below $T_N \approx 5.5$ K, a 3D Néel-type antiferromagnetic order [Fig. 1(b)] is stabilized due to the presence of small perturbative inter-chain couplings [21,23,25,26,27]. In an applied transverse magnetic field along the $a$-axis, the 3D order is suppressed above $B_\perp^{c,3D} = 10$ T [see Fig. 1(b)] [21].

The inter-chain couplings strongly influence the quantum spin dynamics below $T_N$. As illustrated in Fig. 2(a), a spin-flip excitation, which corresponds to $\Delta S = \pm 1$, fractionalizes into two spinons each with a fractional quantum number of spin-1/2. In the Néel-ordered phase, the spinons cannot propagate freely on the chain, but are confined into two-spinon bound states due to the inter-chain couplings. The confining potential increases linearly with the distance between the two spinons [Fig. 2(b)], leading to the discrete levels of spinon-pair bound states, in contrast to spinon continuum of a decoupled chain. Figure 2(c) shows zero-field absorption spectrum of $BaCo_2V_2O_8$ below a strong optical phonon band [23,33]. The absorption spectrum exhibits five sharp peaks with their eigenenergies following a linear dependence on $\zeta_i$ [see Fig. 2(d)], the negative zeros of the Airy function $A_i(-\zeta_i) = 0$, which nicely confirms the confined spinon-pair excitations reported previously [23,25,26,28]. Another important implication of this observation is that the inter-chain couplings provide an effective longitudinal field, which is perturbative and staggered with the peculiar form of $(-1)^i B_{\parallel}$ corresponding to the spin $S_i^z$ on the $i^{th}$ site of the chain. Such a staggered longitudinal field is crucial for the realization of the $E_8$ spectrum, because via the transformation $S_i^z \rightarrow (-1)^i S_i^z$, we can map our antiferromagnetic chain into the ferromagnetic model in Eq. (1).

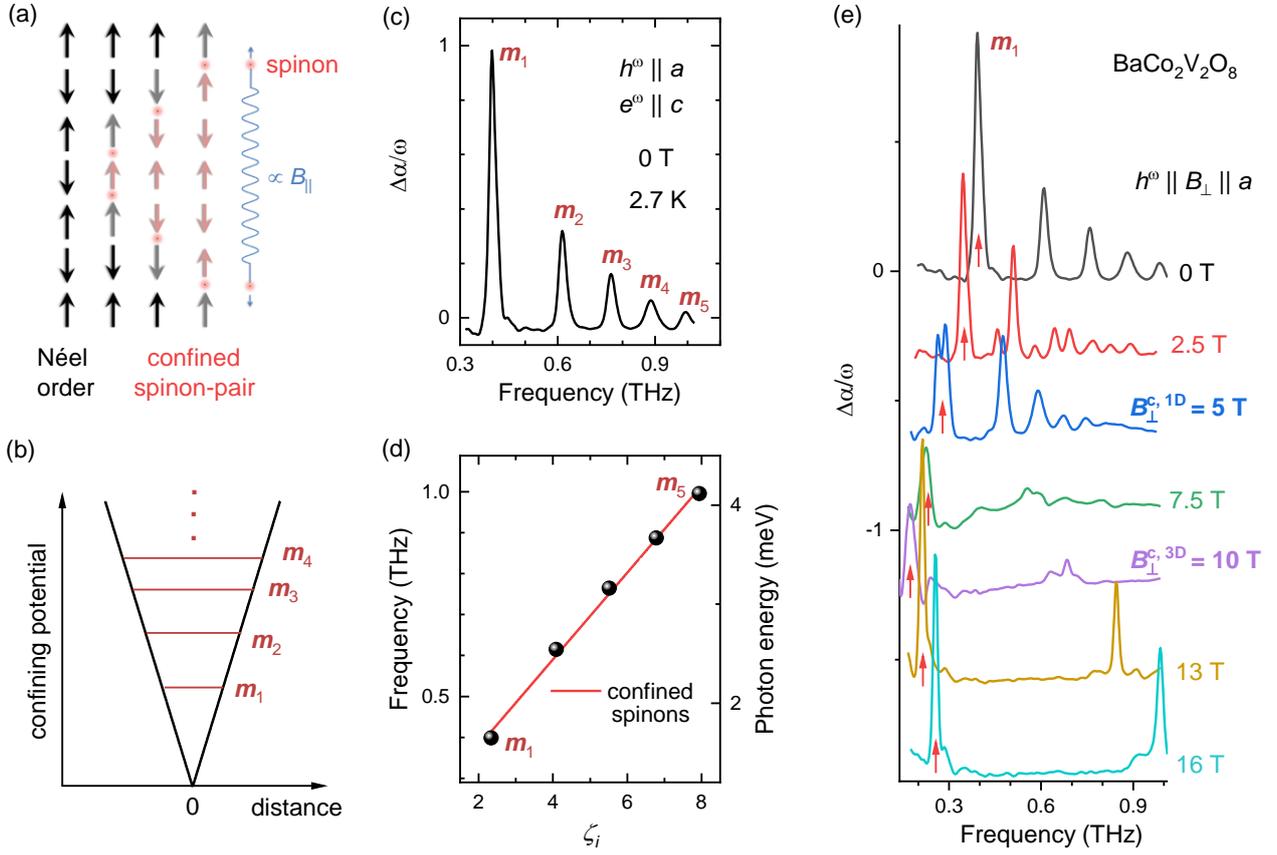

FIG. 2. (a) The inter-chain couplings in the 3D Néel-ordered phase effectively exert a staggered perturbative longitudinal field $(-1)^i B_\parallel$ on the $i^{th}$ spin of the neighboring chain. The two-spinon excitations (red sphere) are confined due to the staggered field. (b) The corresponding confining potential increases linear with the distance between the two spinons, which leads to the discrete levels of two-spinon bound states with energies $m_1, m_2, m_3, \ldots$ (c) Zero-field absorption spectrum exhibits a series of peaks, measured with $h^\omega \parallel a$. $\Delta\alpha$ and $\omega$ denote absorption coefficient and wavenumber, respectively. (d) The energies of the peaks follow a linear dependence on $\zeta_i$, the negative zeros of the Airy function $A_i(-\zeta_i) = 0$, which evidences the confinement of spinons due to the staggered longitudinal fields [23,25,26,28]. (e) Evolution of the absorption spectra in finite transverse fields $B_\perp \parallel h^\omega$. The spectra are shifted vertically by constants proportional to the transverse fields for clarity. The spectra below $B_\perp^{c,3D} = 10$ T in the Néel-ordered phase are complex and exhibit several peaks, while above $B_\perp^{c,3D}$ in the field-induced paramagnetic phase [see Fig. 1(b)] the spectra are mainly featured by two sharp peaks.

While all the aforementioned criteria are found to be fulfilled in BaCo$_2$V$_2$O$_8$ at zero field, it is necessary that they remain fulfilled when applying an external transverse field. In particular, to maintain the collective effects of the staggered fields, the 3D order should not be suppressed before the 1D quantum critical point is reached, i.e. $B_\perp^{c,1D} < B_\perp^{c,3D}$, as illustrated in Fig. 1(b). As we will show below, this condition is indeed realized in BaCo$_2$V$_2$O$_8$.

In a transverse field applied along the crystallographic $a$-axis ($B_\perp \parallel a$), we measured the absorption spectra at 2.7K below $T_N$ with the linearly-polarized THz magnetic field $h^\omega$ along the same orientation (i.e. $h^\omega \parallel B_\perp \parallel a$), see Fig. 2(e) [33]. As indicated by the arrows in Fig. 2(e), the $m_1$ mode observed at 0.4 THz in zero field softens monotonically with increasing field until reaching the minimum frequency of 0.18 THz at 10 T, which is followed by a continuous increase in higher fields (e.g. 0.22 THz at 13 T). The evolution of the lowest-lying mode $m_1$ reflects the field-dependence of the spin excitation gap, which provides the spectroscopic evidence for the suppression of the long-range order above $B_\perp^{c,3D} = 10$ T, consistent with previous thermodynamic measurements [21]. The Néel-ordered and the paramagnetic phases are contrasted by their spin dynamic spectra, which is similar to the behavior reported in an isostructural compound [29]. Below $B_\perp^{c,3D}$, the low-field spectra are characterized by several peaks with different intensities at different energies [33]. In contrast, in the field-induced paramagnetic phase the spectra are dominated by two sharp peaks (e.g. 0.26 and 0.99 THz at 16 T). In addition, a small splitting of the $m_1$ peak (about 0.1 meV) is resolved above 5 T but disappears above 10T, indicating the existence of a weak orthorhombic $ab$-plane anisotropy in the 3D ordered phase [21,30], while no splitting of the higher-energy peaks can be resolved.

The transverse-field dependence of the spin dynamics in the Ising chain systems has been the subject of previous reports, based on experimental studies and/or on numerical simulations, see e.g. Ref. [27,29,31]. Here, we focus on the discussion of the $E_8$ dynamics that was predicted to emerge only in the vicinity of the transverse-field Ising quantum critical point. Figure 3(a) shows the energy ratios of the higher-frequency excitations with respect to the corresponding $m_1$ mode at each field. With increasing field, we observe a continuous increase of all the ratios, and at 5 T, they simultaneously reach the expected values for $m_2$, $m_3$ and $2m_1$, $m_4$, $m_1+m_2$, and $m_5$ of the $E_8$ dynamic spectrum (see Table 1), as indicated by the dashed and dotted lines. Above 5 T, the ratios deviate again from those values of the $E_8$ spectrum. This strongly indicates that we have experimentally realized the $E_8$ spectrum at 5 T [Fig. 3(b)], which also provides the dynamic evidence that $B_\perp^{c,1D} = 5$ T corresponds to the 1D quantum critical field, with the required condition $B_\perp^{c,1D} < B_\perp^{c,3D}$ consistently fulfilled. The value of the critical field agrees with the result of a detailed numerical simulation [31]. At the same time, Fig. 3(b) presents a very crucial feature that the two-particle continua ($2m_1$ and $m_1+m_2$) is characterized by a relatively narrow peak at the onset energies. This observation shows that the continua are not so overwhelming that the higher-energy $E_8$ particles ($m_3$, $m_4$, and $m_5$) can still be resolved, in contrast to the conventional intuitive understanding that the high-energy $E_8$ particles are hidden in a featureless $2m_1$ continuum [32].

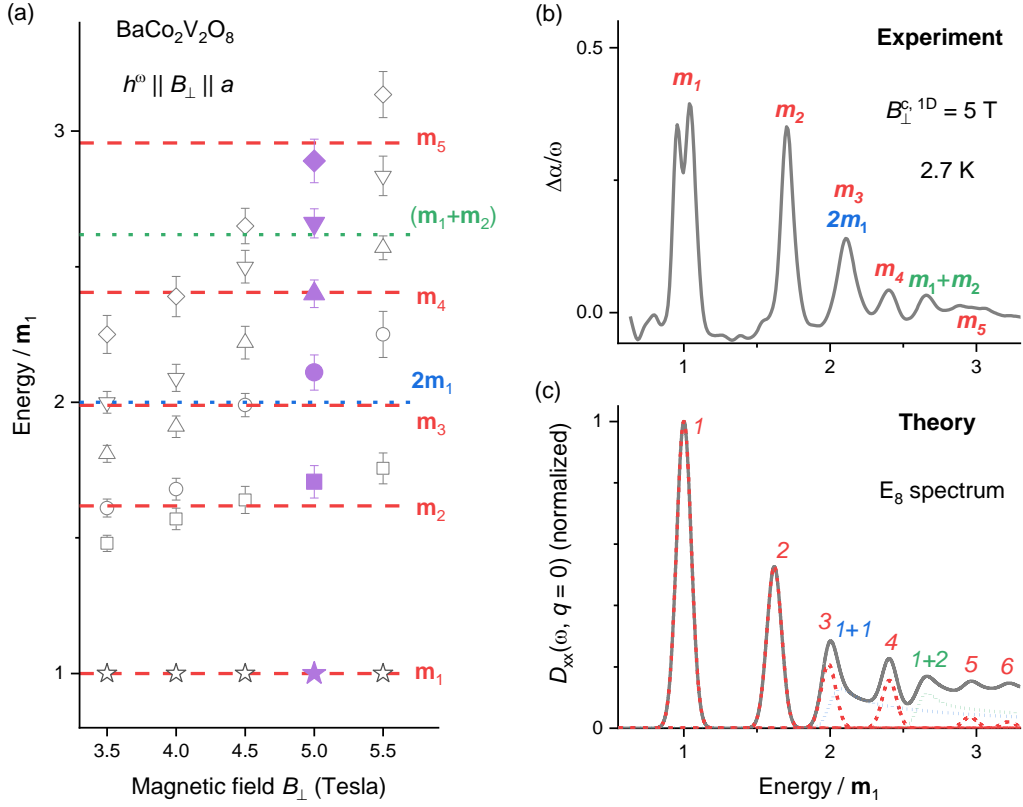

FIG. 3. (a) The ratios of the excitation energies (symbols) increase monotonically as the transverse field approaches 5 T from below, and at 5 T simultaneously reach the theoretically expected values (dashed and dotted lines) for the $E_8$ single- and two-particle excitations $m_2$, $m_3$, $2m_1$, $m_4$, $m_1+m_2$, and $m_5$ (see Table 1). At $B_\perp^{c,1D} = 5$ T, the absorption spectrum in (b) is in excellent agreement with (c) the analytically calculated spin dynamic structure factor $D^{xx}(\omega, q = 0)$ for the quantum integrable model of the $E_8$ dynamic spectrum (solid line) [see Eq. (2)]. The dashed and dotted lines show the separate contributions of the single-particle ($m_1$ to $m_6$) and two-particle ($2m_1$ and $m_1+m_2$) excitations, respectively [33]. Due to a strong phonon band [23], the spectrum in (b) cannot be resolved at higher energies for $m_6$ to $m_8$. The spectra in (c) are broadened with a full-width at half-maximum of $0.1m_1$.

To further elaborate on the dynamic characteristics, we perform analytical calculations of the quantum integrable model of the $E_8$ dynamic spectrum [6,12,13,14]. Corresponding to the transverse THz magnetic field in the Voigt configuration, we calculate the transverse dynamic structure factor

$$D^{xx}(\omega, q = 0) = \sum_{n=0}^{\infty} \int_{-\infty}^{+\infty} \frac{d\theta_1 \cdots d\theta_n}{N(2\pi)^{n-2}} \left| \langle 0|\sigma^x|A_{a_1}(\theta_1)A_{a_2}(\theta_2)\cdots A_{a_n}(\theta_n)\rangle \right|^2 \delta(\omega - E\{a_i\})\delta(P\{a_i\}) \quad (2)$$

for energy transfer $\omega$ and zero momentum transfer $q = 0$, as the wavelength of THz spectroscopy is much greater than the lattice constants of $BaCo_2V_2O_8$. In Eq. (2), $N = \prod_{i=1}^{8} n_i!$, $a_i$ ($i = 1, \cdots, n$) labels a particle with a corresponding mass among $m_1$ to $m_8$, $\theta_i$ is the corresponding rapidity, and $\sigma^x$ is the Pauli matrix associated with the spin component $S^x = \sigma^x/2$. The ground state and the $n$-particle excited state are denoted by $\langle 0|$ and $|A_{a_1}(\theta_1)A_{a_2}(\theta_2)\cdots A_{a_n}(\theta_n)\rangle$, respectively, for $n = 1,2,3\ldots$ The total energy and momentum of the $n$-particle excited state are $E\{a_i\} = \sum_{i=1}^{n} a_i \cosh\theta_i$ and $P\{a_i\} = \sum_{i=1}^{n} a_i \sinh\theta_i$, respectively [33]. In particular, we derive the dynamical response in the two-particle channels $|A_{a_1}(\theta_1)A_{a_2}(\theta_2)\rangle, \{a_1a_2\} = \{m_1m_1, m_1m_2, m_1m_3, m_2m_2\}$ and in the three-particle channel $|A_{a_1}(\theta_1)A_{a_2}(\theta_2)A_{a_3}(\theta_3)\rangle, \{a_1a_2a_3\} = \{m_1m_1m_1\}$ (see Table 1).

The obtained dynamic structure factor is presented in Fig. 3(c) up to the energy of $3.3m_1$ with the peaks broadened by a full-width at half-maximum of $0.1m_1$, which is in accord with the spectral range of our experiment. The lowest energy $m_1$ scaling with the perturbative longitudinal field is set as unit. The contributions of the single-particle excitations ($m_1$ to $m_6$) and of the two-particle continua ($2m_1$ and $m_1+m_2$) are separately plotted as dashed and dotted lines, respectively, while the higher-energy continua with smaller spectral weight are omitted for clarity [33]. The analytical results disclose very peculiar many-body dynamic characteristics. First and foremost, the multi-particle continua are not overwhelming but possess even smaller spectral weight compared with the high-energy $E_8$ single particles. Thus, the single-particle excitations with clearly recognized peaks stand well above the multi-particle continua, profoundly in contrast to the conventional intuitive understanding [32]. Hence, the higher-energy $E_8$ single particles above $2m_1$ should be experimentally resolvable. Furthermore, the multi-particle continuum is not featureless, but exhibits a relatively narrow peak-like maximum just above the onset energy, which is followed by an extended tail. While the peak of $2m_1$ merges coincidently with that of $m_3$ into a single peak and thus cannot be discriminated experimentally, a pronounced peak due to $m_1+m_2$ is clearly discernable. Although the higher-energy multi-particle continua exhibit the similar feature, their spectral weight is very small and hardly recognized in the overall dynamic-structure-factor spectrum [33].

As compared in Fig. 3(b) and 3(c), an overall excellent agreement is achieved between the experimentally observed spectrum at $B_\perp^{c,1D} = 5$ T and the precise dynamic-structure-factor of the $E_8$ dynamics for the single- and two-particle excitations. We emphasize that there are no free-tuning parameters in the field-theory calculation. The agreement between experiment and theory is achieved not only on the energy ratios but also on the relative spectral weights. Although the intensity of the observed $m_1$ peak seems to be relatively low due to the splitting, the ratio of the integrated spectral weight $I_{m2}/I_{m1} \approx 0.61$ is in good agreement with the theoretically predicted value of 0.52 [see Fig. 1(a)].

These results show that the $E_8$ dynamic spectrum is realized in the quasi-one-dimensional antiferromagnetic chain $BaCo_2V_2O_8$ at 5 T, where a 1D transverse-field Ising quantum critical point is evidenced to be hidden under the 3D ordered phase. Our results also imply that the $E_8$ spectrum can generally exist near the quantum critical points of the universality class of the transverse-field Ising chain [31]. The identification of the $E_8$ particles and their multi-particle excitations demonstrates the emergence of the complex symmetry in the vicinity of a quantum critical point and the power of the integrable quantum field theory to describe the complex quantum critical dynamics. Our results in general shed light on the studies of non-equilibrium dynamics in the 1D models [12,18], the quantum simulations in an optical lattice [19], and the deterministic manipulation of quantum many-body states [5].

We thank G. Mussardo and M. Kormos for useful discussions. Z.Z. thanks G. Mussardo for his hospitality at the International School for Advanced Studies (SISSA) during the final stages of this work. J.W. acknowledges additional support from a Shanghai talent program. The work in Tallinn was supported by institutional research funding IUT23-3 of the Estonian Ministry of Education and Research, by European Regional Development Fund Project No. TK134. The work in Cologne was partially supported by the DFG (German Research Foundation) via the project No. 277146847—Collaborative Research Center 1238: Control and Dynamics of Quantum Materials (Subprojects No. A02 and B01).

Z.Z. and K.A. contributed equally to this work.
*Email: wujd@sjtu.edu.cn
†Email: zhewang@ph2.uni-koeln.de

Supplemental Material

# Observation of E₈ particles in an Ising chain antiferromagnet

Zhao Zhang[1], Kirill Amelin[2], Xiao Wang[1], Haiyuan Zou[1], Jiahao Yang[1], Urmas Nagel[2], Toomas Rõõm[2], Tusharkanti Dey[3,4], Agustinus Agung Nugroho[5], Thomas Lorenz[3], Jianda Wu[1], Zhe Wang[3]

[1]Tsung-Dao Lee Institute and School of Physics & Astronomy, Shanghai Jiao Tong University, Shanghai 200240, China
[2]National Institute of Chemical Physics and Biophysics, 12618 Tallinn, Estonia
[3]Institute of Physics II, University of Cologne, 50937 Cologne, Germany
[4]Department of Physics, Indian Institute of Technology (Indian School of Mines), Dhanbad 826004, Jharkhand, India
[5]Faculty of Mathematics and Natural Sciences, Institut Teknologi Bandung, 40132 Bandung, Indonesia

**Theoretical background and analysis.** Integrable quantum field theories are characterized by an infinite number of conserved charges, which dictates scattering processes in one dimension to be elastic and factorizable[1]. Its particle content and interaction rules are encoded in the analytic structure of the *S*-matrix that describes two-body scattering process. In the case that the quasi-particles are non-degenerate with respect to the quantum numbers of the conserved charges, the *S*-matrix is diagonal and its form is uniquely determined by its unitarity and crossing symmetry, once the location of its poles are known. The latter are due to the bound states created in the virtual processes of the two-particle scattering. Treating bound states on the same footing as asymptotic ones, one can extract locations of the poles and restrictions on the spins of the conserved charges from the bootstrap program. Once the mass spectrum is pinned down, we have a basis of the Hilbert space constructed from the Faddeev-Zamolodchikov algebra[2], and linearly dependent vectors are just related by the *S* matrix. The next step is to find the matrix elements of local operators in the basis of out and in states. Upon a crossing symmetry, these are equivalently given by the form factors (FF), which are matrix elements of local observables between vacuum and in states including the antiparticles of out-going particles with negative momentum. The linear dependence among them results in a set of constraints known as Watson's equations involving the *S* matrices, which is solved by minimal two-particle FF's (Ref. [3]). The FF is then the product of the minimal two-particle FF's between two-particle pairs and a symmetric polynomial containing information of the operator and pole structures[4,5].

**Dynamical spin structure factor.** The quantum Ising model in a magnetic field is described by the Hamiltonian

$$H = -J\left(\sum_i \sigma_i^z \sigma_{i+1}^z + h_\perp \sum_i \sigma_i^x + h_\parallel \sum_i \sigma_i^z\right),$$

where $\sigma_i^x$ and $\sigma_i^z$ are the Pauli matrices associated with the spin components $S^\mu = \sigma^\mu/2$ ($\mu = x, z$), and $i$ labels a site position. $h_\perp$ and $h_\parallel$ are the transverse and longitudinal fields, in units of the nearest-neighbor ferromagnetic exchange coupling $J$ between the longitudinal components $S^z$ of the spins. As realized in BaCo₂V₂O₈, the 1D Ising antiferromagnetic model with a staggered longitudinal field $(-1)^i h_\parallel$ can be mapped to this ferromagnetic Ising model by taking the transformation $\sigma_i^z \to (-1)^i \sigma_i^z$. At the quantum critical point $h_\perp = 1$, the continuum limit of this model is described by the $c = 1/2$ conformal field theory universality class perturbed by the primary field $\sigma = \Phi_{(1,2)}$ (Ref.[6])

$$H_{1/2}^{(1,2)} = H_{1/2} + h \int \sigma(x) d^2 x.$$

The operator $\sigma_x$ in the dynamic structure factor (DSF) we compute in the main text corresponds to the conformal family of the energy operator $\epsilon = (1/2, 1/2)$, where the numbers $(1/2, 1/2)$ are its conformal weights[7]. The Hilbert space of this deformed theory is spanned by $|A_{a_1}(\theta_1) \cdots A_{a_n}(\theta_n)\rangle = A_{a_1}^\dagger(\theta_1) \cdots A_{a_n}^\dagger(\theta_n)|0\rangle$, where $A_{a_i}^\dagger(\theta_i)$ is the creation operator of a particle of mass $a_i = m_1, \cdots, m_8$, with rapidity $\theta_i$ (or momentum $p_i = a_i \sinh \theta_i$ and energy $E_i = a_i \cosh \theta_i$) in the Faddeev-Zamolodchikov algebra[2], and $m_1, \cdots, m_8$ are the masses of the eight particles in the model. Since there is no degeneracy in the mass spectrum of $E_8$, particle type is uniquely labeled by its mass. These vectors form a complete basis when we require $\theta_1 > \theta_2 > \cdots > \theta_n$ for in state, or $\theta_1 < \theta_2 < \cdots < \theta_n$ for out states, so $1 = \sum_{n=0}^{\infty} \int_{-\infty}^{\infty} \frac{d\theta_1 \dots d\theta_n}{N(2\pi)^n} |A_{a_1}(\theta_1) \cdots A_{a_n}(\theta_n)\rangle_{in}\,_{in}\langle A_{a_1}(\theta_1) \cdots A_{a_n}(\theta_n)|$, where $n = 1,2,3 \dots$ and $N = \prod_{i=1}^{8} n_i!$ is the symmetric factor accounting for the extension of the integration region using the symmetry of permutation among the same type of particles. By inserting this complete basis in the correlator, the DSF of $\sigma_x$ at zero momentum transfer can be written as

$$D^{xx}(\omega, q=0) = \sum_{n=0}^{\infty} \int_{-\infty}^{+\infty} \frac{d\theta_1 \dots d\theta_n}{N(2\pi)^{n-2}} |\langle 0|\sigma^x|A_{a_1}(\theta_1) \cdots A_{a_n}(\theta_n)\rangle|^2 \delta(\omega - E\{a_i\})\delta(P\{a_i\}),$$

where the FF $F_{a_1 \dots a_n}^{\epsilon}(\theta_1, \dots, \theta_n) = \langle 0|\sigma^x|A_{a_1}(\theta_1) \cdots A_{a_n}(\theta_n)\rangle$ are solved in Ref. [4] to be of the form

$$F_{a_1 \dots a_n}^{\epsilon}(\theta_1, \dots, \theta_n) = \frac{Q_{a_1 \dots a_n}^{\epsilon}(\theta_1, \dots, \theta_n)}{D_{a_1 \dots a_n}(\theta_1, \dots, \theta_n)} \prod_{i<j} F_{a_i a_j}^{min}(\theta_{ij}).$$

The minimal FF $F_{a_i a_j}^{min}(\theta_{ij})$ follows

$$F_{a_i a_j}^{min}(\theta_{ij}) = \left(-i\ \sinh\frac{1}{2}\theta_{ij}\right)^{\delta_{ij}} \prod_{\alpha \in \mathcal{A}_{a_i a_j}} \exp\left[2p_\alpha \int_0^\infty \frac{dt}{t} \frac{\cosh\left(\alpha - \frac{1}{2}\right)t}{\cosh\frac{t}{2}\sinh t} \sin^2\frac{(i\pi - \theta)t}{2\pi}\right]$$

(for the $p_\alpha$'th order pole $\alpha$ in the set of poles $\mathcal{A}_{a_i a_j}$ of the two-body scattering $S$ matrix between $\alpha_i$ and $\alpha_j$), which takes care of the monodromy properties required by the Watsons equations[1], while the symmetric polynomial

$$D_{a_1 \dots a_n}(\theta_1, \dots, \theta_n) = \prod_{\alpha \in \mathcal{A}_{a_i a_j}} \left(\mathcal{P}_\alpha(\theta_{ij})\right)^{i_\alpha} \left(\mathcal{P}_{1-\alpha}(\theta_{ij})\right)^{j_\alpha},$$

with $i_\alpha = n+1$ (resp., $n$), $j_\alpha = n$, if $p_\alpha = 2n+1$ (resp., $2n$), and $\mathcal{P}_\alpha(\theta_{ij}) = (\cos\pi\alpha - \cosh\theta)/2\cos^2(\pi\alpha/2)$, accounts for pole structures. The only operator dependent information is contained in the polynomial $Q_{a_1 \dots a_n}^{\epsilon}(\theta_1, \dots, \theta_n)$, which can be determined from recursive equations for kinematical and bound-state poles and a seed FF of the trace of the energy-momentum tensor. Both the recursive equations and the seed FF are related to the $\sigma$ operator responsible for the perturbation, and subject to symmetry constraints, which can be exploited to pin down the coefficients $c_{a_1 \dots a_n}^k$ in the polynomial $Q_{ab}^{\Phi}(\theta) = \sum_{k=0}^{N_{ab}} c_{ab}^k \cosh^k\theta$. The highly nontrivial task of identifying specific operators out of the linear space of FF's generated by the recursive operations and their kernels was beautifully carried out in Ref. [4,5]. The derived DSF's of various excitations are shown in Supplementary Fig. 1.

**Transmission measurements above and below $T_N$, and in magnetic fields.** Supplementary Fig. 2 presents acquired data of transmission measurements in zero field. As shown in Supplementary Fig. 2(b), transmitted power through the BaCo$_2$V$_2$O$_8$ sample was measured at 10 and 2.7 K above and below the Néel temperature $T_N \approx 5.5$ K, respectively. The ratio of the transmitted power at 10 and 2.7 K is displayed in Supplementary Fig. 2(a), which corresponds to Fig. 2(c) in the main text. The spin excitations **m**$_1$, **m**$_2$, **m**$_3$, **m**$_4$ manifested by peaks in Supplementary Fig. 2(a) can be readily seen in the raw data of 2.7 K in Supplementary Fig. 2(b) as indicated by the arrows, while **m**$_5$ peak becomes evident in the ratio spectrum. The spin excitations observed in the ordered phase disappear as the sample is warmed up above the ordering temperature, but the overall shape of the spectrum remains unchanged. The constant intensity indicates that the reflectivity is insensitive to the small temperature difference in the given frequency range.

At 2.7 K, transmission was measured as a function of an applied transverse field, from which field dependence of the absorption coefficient was derived by taking the 10 K zero-field spectrum as reference. The evolution of the absorption coefficient due to the spin excitations in the applied magnetic field is presented in Supplementary Fig. 3.

**Representation of comparison between experiment and theory.** As presented in Supplementary Fig. 4, the 5 T spectrum of Fig. 3(b) is overlaid on top of the theoretically predicted E$_8$ spectrum of Fig. 3(c). This plot just serves as an alternative representation to show the agreement between the experimental and theoretical results.

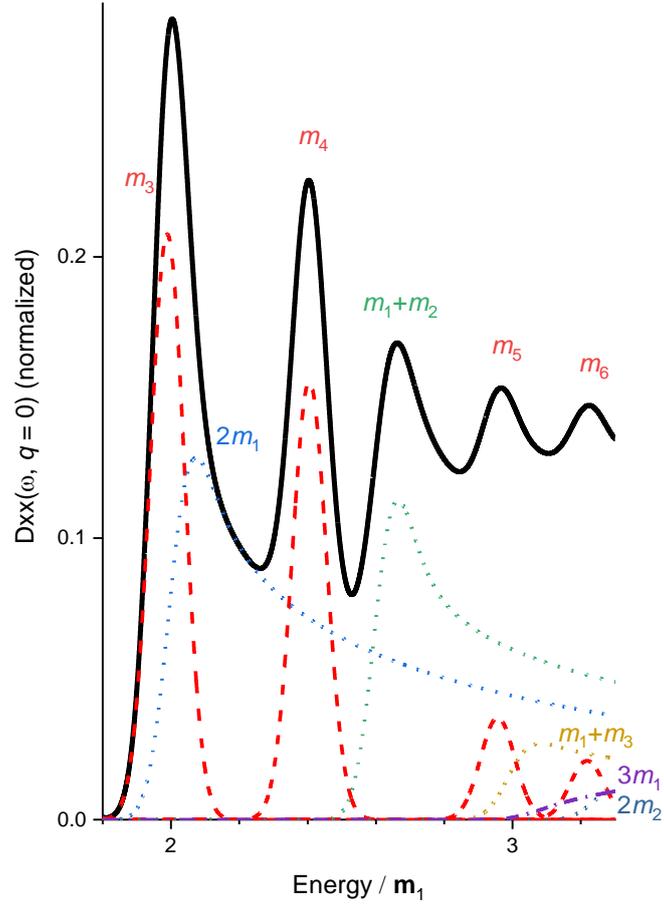

**Supplementary FIG 1.** Analytically calculated spin dynamic structure factor $D^{xx}(\omega, q = 0)$ of the quantum integrable model with the $E_8$ spectrum (solid line) [see Table 1 and Eq. (2)]. The dashed, dotted, and dash-dotted lines show the separate contributions of the single-particle ($m_3$ to $m_6$), the two-particle ($2m_1$, $m_1+m_2$, $m_1+m_3$, and $2m_2$), and the three-particle ($3m_1$) excitations, respectively. The spectra are Gaussian-broadened with a full-width at half-maximum of $0.1 m_1$. The continua of $m_1+m_3$, $3m_1$, and $2m_2$ are hidden due to the peaks of $m_5$ and $m_6$ and the continua of $2m_1$ and $m_1+m_2$, thus cannot be directly recognized from the overall spectrum.

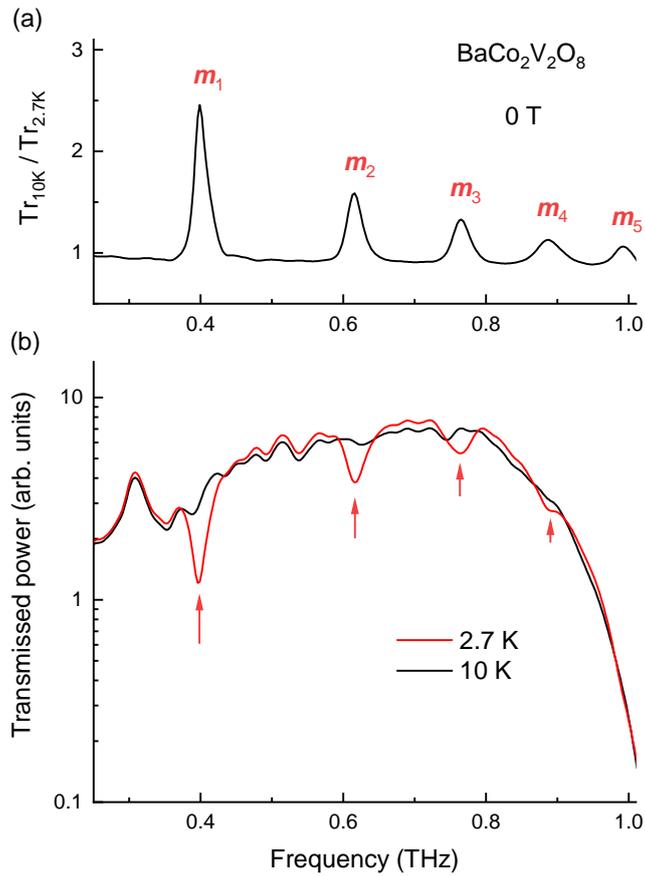

**Supplementary FIG 2.** (a) Ratio of transmitted power measured at 10 and 2.7 K in zero field. The measured power spectra are presented in (b). The spin excitations $m_1$, $m_2$, $m_3$, $m_4$ in (a) can be readily seen in the raw data of 2.7 K in (b) as indicated by the arrows, while $m_5$ peak becomes more evident in the ratio spectrum.

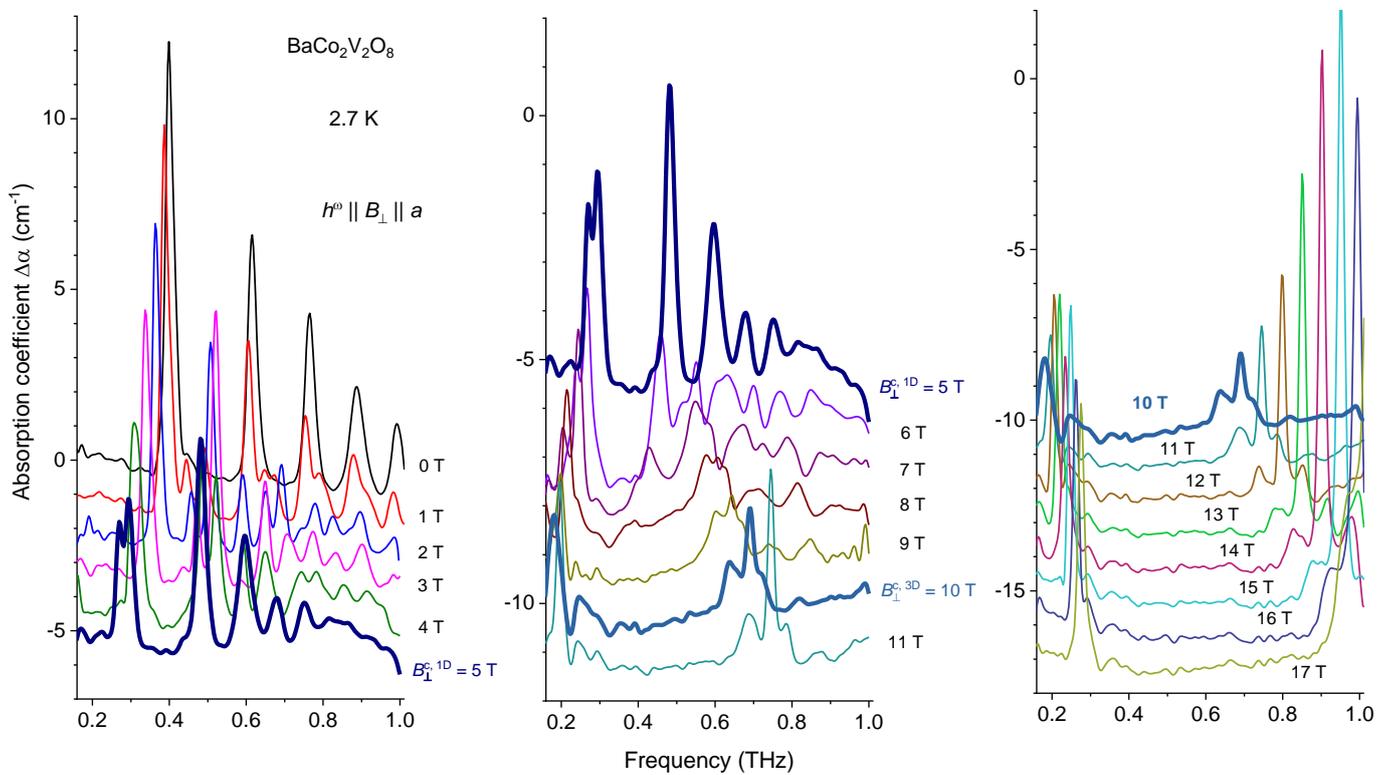

**Supplementary FIG 3.** Absorption coefficient measured at 2.7 K in magnetic fields from 0 to 17 T in 1T step. The spectra are shifted by a constant.

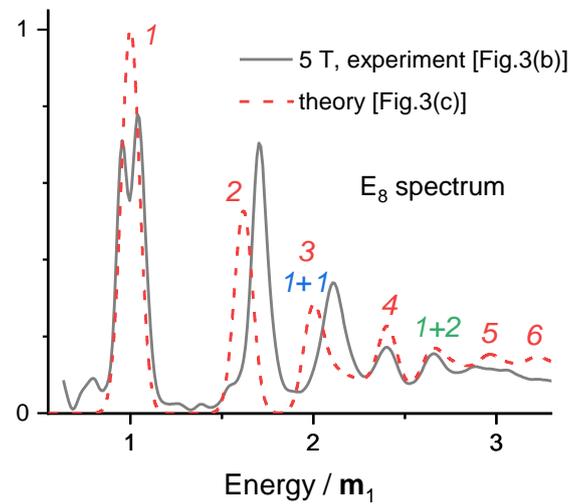

**Supplementary FIG 4.** Overlaying of the 5 T spectrum [Fig. 3(b)] on top of the theoretically predicted $E_8$ spectrum [Fig. 3(c)]. See caption of Fig. 3 for more details.